\documentclass[prd,preprint,nofootinbib]{revtex4}

\usepackage{graphicx}
\usepackage{amssymb}
\usepackage{amsmath}

\begin{document}

%%%%%%%%%new commands added by Takahashi%%%%%%%%%%%%
\renewcommand{\thefootnote}{\#\arabic{footnote}}
\newcommand{\rem}[1]{{\bf [#1]}}
\newcommand{\gsim}{ \mathop{}_ {\textstyle \sim}^{\textstyle >} }
\newcommand{\lsim}{ \mathop{}_ {\textstyle \sim}^{\textstyle <} }
\newcommand{\vev}[1]{ \left\langle {#1}  \right\rangle }
\newcommand{\bear}{\begin{array}}  
\newcommand {\eear}{\end{array}}
\newcommand{\bea}{\begin{eqnarray}}   
\newcommand{\eea}{\end{eqnarray}}
\newcommand{\beq}{\begin{equation}}   
\newcommand{\eeq}{\end{equation}}
\newcommand{\bef}{\begin{figure}}  
\newcommand {\eef}{\end{figure}}
\newcommand{\bec}{\begin{center}} 
\newcommand {\eec}{\end{center}}
\newcommand{\non}{\nonumber}  
\newcommand {\eqn}[1]{\beq {#1}\eeq}
\newcommand{\la}{\left\langle}  
\newcommand{\ra}{\right\rangle}
\newcommand{\ds}{\displaystyle}

\def\SEC#1{Sec.~\ref{#1}}
\def\FIG#1{Fig.~\ref{#1}}
\def\EQ#1{Eq.~(\ref{#1})}
\def\EQS#1{Eqs.~(\ref{#1})}
\def\REF#1{(\ref{#1})}
\def\lrf#1#2{ \left(\frac{#1}{#2}\right)}
\def\lrfp#1#2#3{ \left(\frac{#1}{#2} \right)^{#3}}
\def\GEV#1{10^{#1}{\rm\,GeV}}
\def\MEV#1{10^{#1}{\rm\,MeV}}
\def\KEV#1{10^{#1}{\rm\,keV}}

\def\lrf#1#2{ \left(\frac{#1}{#2}\right)}
\def\lrfp#1#2#3{ \left(\frac{#1}{#2} \right)^{#3}}

\begin{flushright}
IPMU 08-0118
\end{flushright}

\title{
Gravitational Dark Matter Decay and the ATIC/PPB-BETS Excess
}

\author{Fuminobu Takahashi$^{(a)}$ and Eiichiro Komatsu$^{(a,b)}$}

\affiliation{%
$^a$ Institute for the Physics and Mathematics of the Universe, 
     University of Tokyo, Chiba 277-8568, Japan\\
$^b$ Department of Astronomy, The University of Texas at Austin, Austin, TX 78712, USA
}

\date{\today}

\begin{abstract}
  The hidden scalar field, which couples to the visible sector only through
  Planck-suppressed interactions, 
  is a candidate for dark matter owing to its long lifetime. 
  Decay of such a scalar field offers observational tests of this scenario.
  We show that decay of the hidden scalar field can explain the observed
  excess of high-energy positrons/electrons observed by ATIC/PPB-BETS,
  for a suitable choice of the mass and the vacuum expectation value of
  the field. We also show that the same choice of the parameters
  gives the observed dark matter abundance.
  Such a remarkable coincidence suggests that the
  Planck-suppressed interactions may be responsible for the observed excess in
  the cosmic-ray positrons/electrons.
\end{abstract}


\maketitle

The presence of dark matter has been  established firmly by numerous
observations, e.g.,~\cite{Komatsu:2008hk}. While we have not yet
understood the nature of dark matter, recent experimental
data from the cosmic-ray physics may be providing us with new insight
into dark matter properties.
The PAMELA data~\cite{Adriani:2008zr} showed that the positron
fraction starts to deviate from a theoretically expected value for
secondary positrons at around $10$ GeV, and continues to increase up to
about $100$\,GeV. The ATIC balloon-borne experiment
collaboration~\cite{ATIC-new} has recently released their data, showing
a clear excess in the total flux of electrons plus positrons peaked
at around $600 - 700$\,GeV, in agreement with the PPB-BETS
observation~\cite{Torii:2008xu}.  The excess may be explained by
astrophysical sources such as
pulsars~\cite{Aharonian(1995),Hooper:2008kg},
microquasars~\cite{Heinz:2002qj} and gamma-ray bursts~\cite{Ioka:2008cv}.  
An alternative explanation is the decay or annihilation of dark matter particles.

In this letter we focus on the decaying dark matter scenario as an
explanation for the observed excess of high-energy positrons and
electrons.  In order to explain the ATIC/PPB-BETS excess, the dark matter particles
must produce electrons and positrons with a hard energy spectrum, and
satisfy the following properties:
\bea
\label{mass}
m &\simeq& (1 - 2) {\rm\,TeV},\\
\tau  & = &{\cal O}(10^{26}){\rm\, sec},
\label{lifetime}
\eea
where $m $ and $\tau$ are the mass and the lifetime of the dark
matter particles, respectively.  
The constraint on $m$ comes from the observed energy spectrum of
the positron/electron excess, which has been detected up to $\sim$\,TeV energy
with a suggestive cut-off at $\sim 600$~GeV~\footnote{
Such an excess is indicative of 
a particle with $m\sim 1.2$~TeV for the two-body decay, and $m\sim 1.8$~TeV for the three-body
decay.}, while that on $\tau$ comes from the observed flux.
We assume that the decaying dark matter accounts for most of the observed dark matter 
density throughout this letter.

We shall show that, if the dark matter is a hidden scalar field~\footnote{
See, e.g., Refs.~\cite{gravitino,hidden-gauge,composite,
decaying-DM-sonota,Arvanitaki:2008hq} for other explanations.
} whose
decay through Planck-suppressed dimension $6$ operators explains the
ATIC/PPB-BETS excess, the observed dark matter abundance can also be
explained naturally and simultaneously. This is a non-trivial
coincidence; thus,  
we shall conclude that the dark matter
decaying through gravitational (Planck-suppressed) interactions may be
responsible for the observed excess in the cosmic-ray positron/electron
flux. 

\vspace{5mm}

The longevity of dark matter is a puzzle, especially if its mass is as
heavy as $1$\,TeV; the dark matter
particle may sequester itself from the standard-model sector, or it may be
protected by some discrete symmetry, or perhaps both. 

Let us assume that the dark matter particle is a scalar field of  mass
given by \REF{mass}, and
has only Planck-suppressed interactions with the standard-model particles.  
If the scalar is a singlet under any symmetries, the lifetime will be much
shorter than the present age of the universe, as the scalar field can
certainly have
dimension $5$ operators. Let us therefore assume that the hidden scalar
is charged under a symmetry (say, $Z_2$ symmetry).
Then the hidden scalar field decays through Planck-suppressed 
dimension $6$ operators, and the lifetime can be as long as \REF{lifetime} for
an appropriate amount of the symmetry breaking.

For concreteness, we consider the following form of the Lagrangian
density, the so-called $f(\phi) R$ gravity~\footnote{
Here we adopt a convention that $\sqrt{-g}$ is included in the Lagrangian density,
following Ref.~\cite{Watanabe:2006ku}.
}:
\beq
{\cal L} \;=\;  \sqrt{-g} \left( \frac{1}{2} f(\phi) R - \frac{1}{2} \partial_\mu \phi \partial^\mu \phi - V(\phi) \right)
+ {\cal L}_m,
\label{eq:fphiR}
\eeq
where $V(\phi)$ is the scalar potential of $\phi$ and ${\cal L}_m$ is
the matter Lagrangian density.
Note that this particular choice of Lagrangian (i.e., $f(\phi)R$
 gravity) is not essential. Our result applies to any models in which 
 the decay rate is (approximately) given by (\ref{decay-rate}).
As we shall discuss later, our argument applies to
other set-ups, e.g. supergravity theories, in a straightforward
way.
We use the following form of $f(\phi)$,
\beq
f(\phi) \;=\; M_P^2 + \xi \left(\phi^2 -v^2\right), 
\label{eq:fphi}
\eeq
where $M_P \simeq 2.4 \times 10^{18}$\,GeV is the reduced Planck mass,
$\xi$ is a numerical coefficient of order unity, and $v$ denotes the
vacuum expectation value (vev) of $\phi$, i.e., $v \equiv \la
\phi\ra$.  
The scalar potential, $V(\phi)$, is chosen such that  the scalar field,
$\phi$, acquires a vev, $v$~\footnote{  
  A potential production of domain walls can be
  made harmless by making the imposed $Z_2$ symmetry approximate rather
  than exact, e.g., by adding a small explicit breaking of the $Z_2$
  symmetry~\cite{Vilenkin:1981zs}.
  It is even possible that the
  vev of $\phi$ is induced entirely by an explicit breaking of the
  $Z_2$ symmetry; domain walls are not produced in
  this case, and thus they would not affect our arguments.
}.
The conventional Einstein gravity is restored in the
low energy limit, where $\phi$ has settled into the vev, i.e., $f(v)=M_P^2$.
This form of $f(\phi)$ is realized when we impose a $Z_2$
symmetry on $\phi$, which is spontaneously broken by $\la \phi \ra$ in
the vacuum.

Let us decompose $\phi$ into the classical part ($v$) and the quantum
fluctuation ($\sigma$) as
%%
%\beq
$\phi \;=\; v + \sigma$,
%\eeq
%%
and assume that $v\neq 0$.
The previous work \cite{Endo:2006qk,Watanabe:2006ku} has shown that 
$\sigma$ is generically coupled to any matter fields, even if
$\phi$ does not have direct couplings with them in ${\cal
  L}_m$, as long as the matter fields are not conformally invariant.\footnote{
  The scalar field can also decay into gauge fields, which are
  conformally invariant at the
  tree level -- the conformal invariance is broken at the one-loop
  level~\cite{Endo:2007ih}. 
  The presence of such interactions does not
  change our arguments, as the amplitudes are one-loop suppressed.
}
The interaction vertices
are induced by the mixing of $\sigma$ with gravity.  It may be easier
to understand how the interactions arise by performing the
Weyl transformation
to make the gravity canonically normalized (i.e., the Einstein
gravity). Since the Weyl transformation depends on the scalar $\phi$,
the interactions between $\sigma$ and the matter fields are induced in
the Einstein frame. See \cite{Watanabe:2006ku} for details. 

As for the matter Lagrangian density, ${\cal L}_m$, let us consider  another
scalar field, $\chi$, with the following form for simplicity:
\beq
{\cal L}_m \;=\;  -\frac{\sqrt{-g}}{2} \left(\partial_\mu \chi \partial^\mu \chi + m_\chi^2 \chi^2 \right).
\eeq
The hidden scalar, $\sigma$, then decays into a pair of $\chi$'s through 
vertices $\sim v \sigma (\partial \chi)^2 / M_P^2$ and 
$v m_\chi^2\sigma \chi^2 / M_P^2$. 
 The decay rate has
been calculated by Ref.~\cite{Watanabe:2006ku} for a general form of
$f(\phi)$ and is given by
\beq
\Gamma \;=\;  \frac{\hat{g}_\chi^2}{8 \pi m_\sigma} \left(1-\frac{4 m_\chi^2}{m_\sigma^2}\right)^\frac{1}{2},
\eeq
where
\beq
\hat{g}_\chi \;\equiv\; \frac{f^\prime(v)}{2 M_P^2} \frac{m_\chi^2+\frac{m_\sigma^2}{2}}{\sqrt{1+\frac{3}{2} (\frac{f^\prime(v)}{M_P})^2}}.
\eeq
Here, $m_\sigma$ is the mass of $\sigma$, and 
$f^\prime(v)\equiv \left.\partial f/\partial\phi\right|_{\phi=v}$.  
If $m_\chi$ is much smaller than
$m_\sigma$, the decay rate is approximately given by
\beq
\Gamma \;\simeq\;  \frac{\xi}{32 \pi} \lrfp{v}{M_P}{2} \frac{m_\sigma^3}{M_P^2},
\label{decay-rate}
\eeq
where we have used the form of $f(\phi)$ given by \EQ{eq:fphi}, and 
assumed $v \ll M_P$ (which gives $f^\prime(v)\ll M_P$).

\begin{figure}[t]
\includegraphics[scale=0.5]{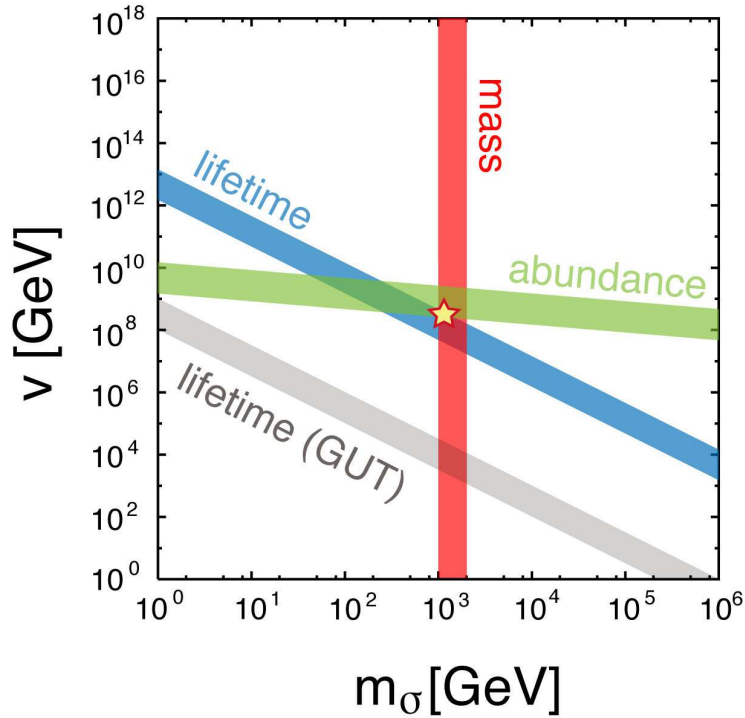}
\caption{Constraints on the mass, $m_\sigma$, and the vacuum expectation
 value, $v$, of the hidden scalar field dark matter from the ATIC/PPB-BETS
excess and the cosmological dark matter abundance. The energy and flux
 of  cosmic-ray positrons/electrons detected by ATIC/PPB-BETS give
 $m_\sigma$ (vertical (red) band) and the lifetime (oblique (blue)
 band), respectively. The nearly horizontal (green) band shows the
 abundance constraint. The widths of the bands show the uncertainties in
 the details of the model: for the mass and lifetime we vary $\xi N$
 from 0.1 to 10, and for the abundance we vary $A$ from 0.1 to 10. 
Three constraints  meet at
 one point  represented by a star. The lower (gray) band
 shows the constraint from the lifetime with the Planck-suppressed
 interaction (\ref{decay-rate}) replaced by the GUT-scale-suppressed
 one, $M_{\rm GUT} = 2 \times 10^{16}$\,GeV.
}
\label{fig:coin}
\end{figure}

In our scenario the scalar $\phi$ is the dominant
component of dark matter, which decays into the standard-model
particles through the gravitational couplings, and 
the decay products are the source for the observed
excess in the cosmic-ray positrons/electrons.  This may be realized if the
scalar matter field, $\chi$, promptly decays into an electron-positron pair~\cite{Cholis:2008vb}\footnote{
The energy spectrum of the electrons and positrons depends on the details of
their production. It may be possible to distinguish different production processes by
measuring the spectrum precisely in future observations~\cite{Chen:2008fx}.
}.
 In order to meet the required lifetime \REF{lifetime}, the vev of $\phi$ must satisfy
\beq
v \;\simeq\; 2 \times 10^8 {\rm \,GeV} \,(\xi N)^{-\frac{1}{2}} 
\lrfp{m_\sigma}{1{\rm\, TeV}}{-\frac{3}{2}} \lrfp{\tau_\phi}{10^{26}{\rm\, sec}}{-\frac{1}{2}},
\label{vev}
\eeq
where we have assumed that $\sigma$ decays into different $N$ scalars, $\chi_i$ $(i = 1,\cdots N)$.
Thus, in this model, the mass and vev of $\phi$ are fixed by the requirements
(\ref{mass}) and (\ref{lifetime}). In Fig.\ref{fig:coin} we show 
(\ref{mass}) and (\ref{vev})  on the $(m_\sigma, v)$-plane,
where we have varied $\xi N$ from $0.1$ to  $10$.

So far, we have merely shown that we can explain two observables 
(energy and flux of the high-energy cosmic ray positrons and electrons)
by  tuning two parameters, $m_\sigma$ and $v$, which may not be so
remarkable. In the following we shall show that the {\it same} set of
parameters can explain the cosmological abundance of $\phi$
simultaneously, i.e., two parameters can explain three observables. 

We estimate the cosmological abundance of $\phi$ as follows.  In general we
expect that the initial position of $\phi$ set 
during inflation was different from the potential minimum in the low
energy.  For instance, if the $Z_2$ symmetry was respected during inflation,
the $\phi$ field would sit at the origin, displaced from the low 
energy minimum by $v$.
The $\phi$ field would start to oscillate about the potential
minimum when the Hubble parameter became comparable to the mass,
i.e., $H \simeq m_\phi$, with an amplitude around $v$, where $m_\phi$
is the mass at the initial position.  
Throughout this letter we shall assume $m_\phi \simeq
m_\sigma$ only for simplicity: $m_\phi$ can be different from $m_\sigma$
in general, depending upon the shape of $V(\phi)$.
If $m_\sigma\neq m_\phi$, it is $m_\sigma$ that must satisfy
the mass constraint given by \REF{mass}.

Assuming that reheating of the universe after inflation has been
completed by the beginning of oscillations of $\phi$ (see below for the
other case), we estimate the
cosmological abundance of $\phi$ as
\bea
\frac{\rho_\phi}{s} &=& 
\frac{\frac{A}{2}\, m_\phi^2 v^2}{\frac{2 \pi^2 g_*}{45} \,(3 m_\phi^2 M_P^2 \frac{30}{\pi^2 g_*})^\frac{3}{4}},\non\\
&\simeq& 6 \times 10^{-10}A~ {\rm GeV}\lrfp{g_*}{100}{-\frac{1}{4}} \lrfp{v}{10^9 {\rm \,GeV}}{2} \lrfp{m_\sigma}{1{\rm\,TeV}}{\frac{1}{2}},
\eea
where 
$\rho_\phi$ is the energy density of $\phi$, $s$ the entropy
density, and $g_*$ the relativistic degrees of freedom at $H = m_\phi$.
We have introduced a numerical coefficient
$A$ 
to parametrize an ${\cal O}(1)$ uncertainty in the above
estimate. 
Note that we have used $m_\phi\simeq m_\sigma$.
One may also write this result in
the following form:
\beq
\Omega_\phi h^2 = 0.2A \lrfp{g_*}{100}{-\frac{1}{4}} \lrfp{v}{10^9 {\rm \,GeV}}{2} \lrfp{m_\sigma}{1{\rm\,TeV}}{\frac{1}{2}}.
\label{eq:om}
\eeq
Here, $\Omega_\phi$ is the density parameter of $\phi$, and $h$ the present
Hubble in units of 100 km/s/Mpc.  
The predicted dark matter abundance, for $v$ and $m_\sigma$ that are
required to  explain the ATIC/PPB-BETS excess, is in remarkable agreement 
with the measured dark matter abundance,
$\Omega_{m}h^2 \simeq 0.11$~\cite{Komatsu:2008hk}.  No tuning of
parameters, apart from choosing the two parameters, $v$ \REF{vev} and
$m_\sigma$ \REF{mass}, to explain the ATIC/PPB-BETS excess, was required.
We have plotted
the region of $(m_\sigma, v)$ where the dark matter abundance
agrees with the observed value in Fig.~\ref{fig:coin}. In the
figure we have varied $A$ from $0.1$ to $10$.

The three conditions, the mass (\ref{mass}), the lifetime
(\ref{lifetime}), and the cosmological abundance (\ref{eq:om}) are a
priori independent of one another. Nevertheless, if we assume that the hidden scalar
dark matter is coupled to the visible sector only by the
Planck-suppressed interactions, those three conditions meet at 
a single point on the $(m_\sigma, v)$ plane,
i.e., $m = {\cal O}(1)$\,TeV and $v = {\cal O}(10^9)$\,GeV. 
Did this happen merely by chance? Such a remarkable coincidence may suggest
that the Planck-scale physics is 
playing an important role in the decaying dark matter scenario that
accounts for the ATIC/PPB-BETS excess. For comparison,  we also show in Fig.~\ref{fig:coin} 
a constraint from the lifetime if the cut-off scale is the grand unification theory (GUT) scale instead of the Planck-scale
(the lower (gray) band). The other two lines (from mass and
abundance) are the same. The three lines no longer meet at one point.

\vspace{5mm}
To obtain the abundance \REF{eq:om} we have assumed that the reheating
has been completed before the $\phi$ field began to oscillate. This assumption
can be translated into the lower bound on the reheating temperature:
$T_R \gtrsim \sqrt{m_\phi M_P} \sim 10^{10}$\,GeV.  
On the other hand, if the reheating temperature was as high as $10^{14}$
GeV, the thermal production ($\chi 
\chi \rightarrow \phi \phi$) through the Planck-suppressed interaction,
$\sim \phi^2 \chi^2/M_P^2$ (see Sec.~IV of Ref.~\cite{Watanabe:2007tf}),
would give a significant contribution to the
dark matter abundance, while the thermal production can be neglected for $T_R <
10^{14}\,$GeV. Therefore the above estimate \REF{eq:om} is valid for
the reheating temperature between $10^{10}$\,GeV and $10^{14}$\,GeV.

What if the reheating was not completed when the $\phi$ began to oscillate,
i.e., $T_R \lesssim 10^{10}$\,GeV?  The abundance would be diluted by the
entropy production during reheating, and $\Omega_\phi h^2$ would be given by
 \beq
 \Omega_\phi h^2 \;\simeq\; 0.06\, A \lrfp{v}{10^9{\rm GeV}}{2} \lrf{T_R}{10^9{\rm GeV}}.
 %0.0577
 \eeq
 which is necessarily smaller than the previous estimate
 (\ref{eq:om}), and therefore the agreement of three lines 
shown in Fig.~\ref{fig:coin} would not be as good. 

Our discussion so far did not use supergravity; however, in supergravity there is a notorious gravitino problem~\cite{Weinberg:zq,Krauss:1983ik,
 Kawasaki:2006gs} (see \cite{Kawasaki:2008qe} and references therein
 for the recent constraint from Big Bang Nucleosynthesis), which
 requires care when the reheating temperature is higher than
 $10^{10}$\,GeV or so. 
The reheating temperature higher than $10^{10}$\,GeV is allowed
for a gravitino mass in the following ranges: (i) $m_{3/2} \lesssim 10$\,eV~\cite{Viel:2005qj},
(ii) $m_{3/2} \gtrsim 100$\,GeV (if the gravitino is the lightest supersymmetric particle)
and (iii) $m_{3/2} \gtrsim {\cal O}(10)$\,TeV. The presence of $R$-parity violation and/or
a light $R$-parity odd field in a hidden sector may enlarge the allowed parameter space in some cases, 
but further discussion is beyond the scope of this letter.

 As mentioned earlier, a coupling similar to (\ref{eq:fphiR}) is generically present in the supergravity. 
 In the conformal frame there is a term given by
 \beq
 {\cal L} =  -\frac{\sqrt{-g}}{2} e^{-K/3} R + \cdots,
 \eeq
 where $K$ is the K\"ahler potential.
 After performing the field-dependent Weyl transformation, we
 generically obtain quartic couplings such as
  $\sim \int d^4 \theta\, |\Phi|^2 |Q|^2/M_P^2$ in the Einstein frame,
 where  $\Phi$ and $Q$ denote the dark matter and a matter field, respectively. 
 We assume that  $\Phi$  is  odd under a $Z_2$ symmetry, and the lowest component, $\phi$, is the hidden scalar dark matter.
 The decay into the fermionic partner and the gravitino is assumed to be kinematically forbidden.
 The scalar matter field that appeared in our discussion so far, $\chi$,
 may be identified with a slepton within the context of supersymmetry.  The decay into a pair of sleptons is induced
by a quartic coupling in the K\"ahler potential such as
 $\sim \int d^4 \theta\, |\Phi|^2 |e_i|^2/M_P^2$,
 where $e_i$ denotes the right-handed lepton superfield in the $i$-th generation.
The decay into a pair of sleptons through this coupling is suppressed by
 $(m_{\tilde e_R,i}/m_\phi)^4$ with respect to (\ref{decay-rate}), if we redefine the vev 
as $v \equiv \la |\phi|\ra/\sqrt{2}$~\footnote{Note that $\phi$ is a complex scalar here.}.
 However,  as the suppression is not so significant for the slepton mass
 of ${\cal  O}(100)$\,GeV, our previous arguments are still valid without
 modification. 
 For instance, the suppression factor including the phase space is $\sim 0.02$ for
 $m_\phi = 1.4{\rm TeV}$ and $m_{\tilde e_R,i} = 600{\rm GeV}$. 
 We have varied the decay rate by two orders of magnitudes in Fig.~\ref{fig:coin},
 which can account for this kind of possible uncertainty. In fact, the
 agreement of the three lines in Fig.~\ref{fig:coin} becomes even better
 in this case.

If the $Z_2$ symmetry is explicitly broken by a small amount, $\epsilon \ll
 M_P$,   the vev of $\phi$ is expected to be of order of $\epsilon$. Also there
 may be a linear term in the K\"ahler potential: $\delta K \sim
 \epsilon\, \Phi + {\rm h.c.}$. In the presence of such a linear term,
 the initial position of the $\phi$ field during inflation is naturally
 displaced from the potential minimum by ${\cal
   O}(\epsilon)$~\cite{Ibe:2006am}.  Thus our estimate on the
 cosmological abundance of the $\phi$ is also valid in this case.

\vspace {5mm}

In this letter we have shown that a hidden scalar field dark matter, which 
couples to the standard-model sector only through the
Planck-suppressed dimension $6$ interactions, can explain 
 the  excess of cosmic-ray positrons/electrons observed by 
ATIC/PPB-BETS for a suitable choice of two parameters: the mass and the vacuum
expectation value. We have also shown that the same parameters, without
any further tuning or introduction of parameters, yield the correct dark
matter abundance.   Such a non-trivial
coincidence suggests that  the dark matter decaying through
the Planck-suppressed interactions may be responsible for the ATIC/PPB-BETS
excess. We have presented an explicit example using the so-called
$f(\phi)R$ gravity, 
and also shown how it can be  embedded in supergravity easily.
We have seen that the vev of the hidden scalar field is necessarily
${\cal O}(10^{8-9})$\,GeV in order to account for the ATIC/PPB-BETS
excess. The origin of such an intermediate  
scale would require further explanation.

%%%%%%%%%%%%%%%%%%%%
%%%%%%%%%%%%%%%%%%%%

%%%%%%%%%%%%%%%%%%%%%%%%%%%%%%%%%%%%%%%%%%%%
\begin{acknowledgements}
%%%%%%%%%%%%%%%%%%%%%%%%%%%%%%%%%%%%%%%%%%%%
F.T. thanks K. Nakayama for comments.
This work is supported by World Premier International Research Center
 Initiative (WPI Initiative), MEXT, Japan, and by NSF grant PHY-0758153.
E.K. acknowledges support from the Alfred P. Sloan Research Foundation.

 %%%%%%%%%%%%%%%%%%%%%%%%%%%%%%%%%%%%%%%%%%%%
\end{acknowledgements}
%%%%%%%%%%%%%%%%%%%%%%%%%%%%%%%%%%%%%%%%%%%%

%%%%%%%%%%%%%%%%%%%%%%%%%%%%%%%%%%%%%%%%%%%%

%%%%%%%%%%%%%%%%%%%%%%%%%%%%%%%%%%%%%%%%%%%%

\end{document}